\documentclass[a4paper,11pt]{article}
\pdfoutput=1 

\usepackage{jinstpub} 
\usepackage[utf8]{inputenc} 
\usepackage{comment}

\title{Single photon time resolution of photodetectors at high rate: Hamamatsu R13742 MaPMT and R10754 MCP-PMT}

\author[a,b]{M. Calvi,}
\author[a,b]{S. Capelli,}
\author[a,b]{P. Carniti,}
\author[b]{C. Gotti}
\author[b]{and G. Pessina}

\affiliation[a]{Università di Milano-Bicocca, Dipartimento di Fisica G. Occhialini, Piazza della Scienza 3, 20126 Italy}
\affiliation[b]{INFN Sezione di Milano-Bicocca, Piazza della Scienza 3, 20126 Italy}

\abstract{This paper reports on the time resolution of two photodetectors operated as single photon counters at high rate: a Hamamatsu R13742-103-M64 ``conventional’’ (based on metal dynodes) multi-anode photomultiplier tube (MaPMT) and a Hamamatsu R10754-07-M16 microchannel plate photomultiplier tube (MCP-PMT).
The MCP-PMT shows a time resolution (transit time spread, or jitter) of $\sim$70~ps FWHM ($\sim$30~ps RMS) at low photon rates, but saturates above $\sim$100~kHz/mm$^2$.
The MaPMT can handle photon counting rates up to the highest tested, 10~MHz/mm$^2$.
Its time resolution is $\sim$250~ps~FWHM ($\sim$110~ps RMS) when only the pixel center is illuminated, but pixel edge effects degrade the resolution to $\sim$400~ps~FWHM ($\sim$170~ps RMS) when the entire pixel area is illuminated.}

\begin{document}
\maketitle
\flushbottom

\section{Introduction}
\label{sec:intro}

Technologies that enable detection of quanta of electromagnetic radiation, i.e. single photons, are widely used in particle physics, and find application in the fields of medical imaging, LIDAR, quantum science, biology, and others \cite{Review1, Review2}.
Despite their bulkiness, high operating voltage and other drawbacks, vacuum-based detectors still outperform silicon-based detectors when it comes to single photon sensitivity, as the dark count rate at room temperature is several orders of magnitude lower.
The difference is further enhanced if the devices need to operate in presence of radiation. The high sensitivity to displacement damage of silicon photomultipliers (SiPM) makes their use in accelerator environment challenging, and even more so if sensitivity to single photons is required \cite{SiPMreview,SiPMnostro}.
This is the case of next generation ring imaging Cherenkov (RICH) detectors, used for particle identification in high energy physics experiments, and in particular in the LHCb experiment at the large hadron collider (LHC) operating at CERN \cite{LHCbRICH}.
The increasing luminosity of the accelerator and experiments calls for detectors able to operate at high photon counting rate.
The use of precise timing information has been proposed to improve particle identification performance \cite{LHCbUpgradeII, LHCbUpgradeIIp}.

In vacuum-based photodetectors, a photon entering the device causes the emission of a photoelectron from the photocathode, with typical quantum efficiency up to $\sim$30\%.
The photoelectron enters an electron multiplier stage with gain of $\sim$10$^6$, and a charge signal of $\sim$100~fC is collected at the anode.
Electron multiplication in a ``standard'' PMT is provided by a series of discrete electrodes, or dynodes.
Each stage provides a gain of 3-5 depending on the accelerating voltage and dynode material.
In this work we used the multi-anode PMT (MaPMT) Hamamatsu R13742-103-M64, with 12 metal dynode stages\footnote{Serial number: FA0047. The R13742 a device equivalent to the R11265, produced by Hamamatsu for the LHCb RICH detectors.}.
The charge is collected on $2.88 \times 2.88$~mm$^2$ square anodes. The device has $8\times 8$ pixels in a total area of $26.2 \times 26.2$~mm$^2$, and a total active area ratio of 78\%.
The maximum bias voltage is 1100~V.
We used the recommended voltage divider ratio of 2.3:1.2:1:...:1:0.5.
The average gain of the tested sample is $1.5 \times 10^6$ at 900~V, $4.5 \times 10^6$ at 1000~V.
The single photon counting performance of this PMT model at low photon rate was characterized in a previous work, time resolution excluded \cite{R11265}.

An alternative structure for electron multiplication in a vacuum tube is the microchannel plate (MCP).
In this case, multiplication occurs in $\sim$10~$\mu$m diameter channels etched in a mm-thick slab of a high resistivity material (typically lead glass).
Two MCP slabs are usually stacked in a Chevron (v-shape) configuration to reach a gain of $\sim$10$^6$.
The shorter path of the electron cloud in the microchannels compared to the case of discrete dynodes translates to a smaller transit time spread, hence a better time resolution compared to conventional PMTs \cite{ReviewThierry}.
However, due to the high resistivity of the base material, each microchannel has a $\sim$ms recharge time after each signal, which leads to saturation at high rate \cite{MCPsat1,MCPsat2}.
In this work we measured the Hamamatsu R10754-07-M16 MCP-PMT, a 2-stage MCP with 10~$\mu$m diameter channels\footnote{Serial number: KT0862. The R10754 MCP-PMT was developed for the TOP detector of the Belle experiment \cite{BelleTDR}.}.
The anodes are arranged in $4 \times 4$ square pixels of $5.28 \times 5.28$~mm$^2$, a total area of $27.6 \times 27.6$~mm$^2$ and an active area ratio of 58\%.
The maximum bias voltage of the device is 2300~V. We used the recommended voltage divider ratio (0.5:2.5:2.5:2.5:1). The gain is $1 \times 10^6$ at 2150~V and $3.5 \times 10^6$ at 2250~V.
We compared its performance with that of the R13742, focusing on timing resolution (transit time spread) and its dependence on photon rate.

\section{Test setup and analysis method}

\begin{figure}[t]
\centering
\includegraphics[width=.6\textwidth]{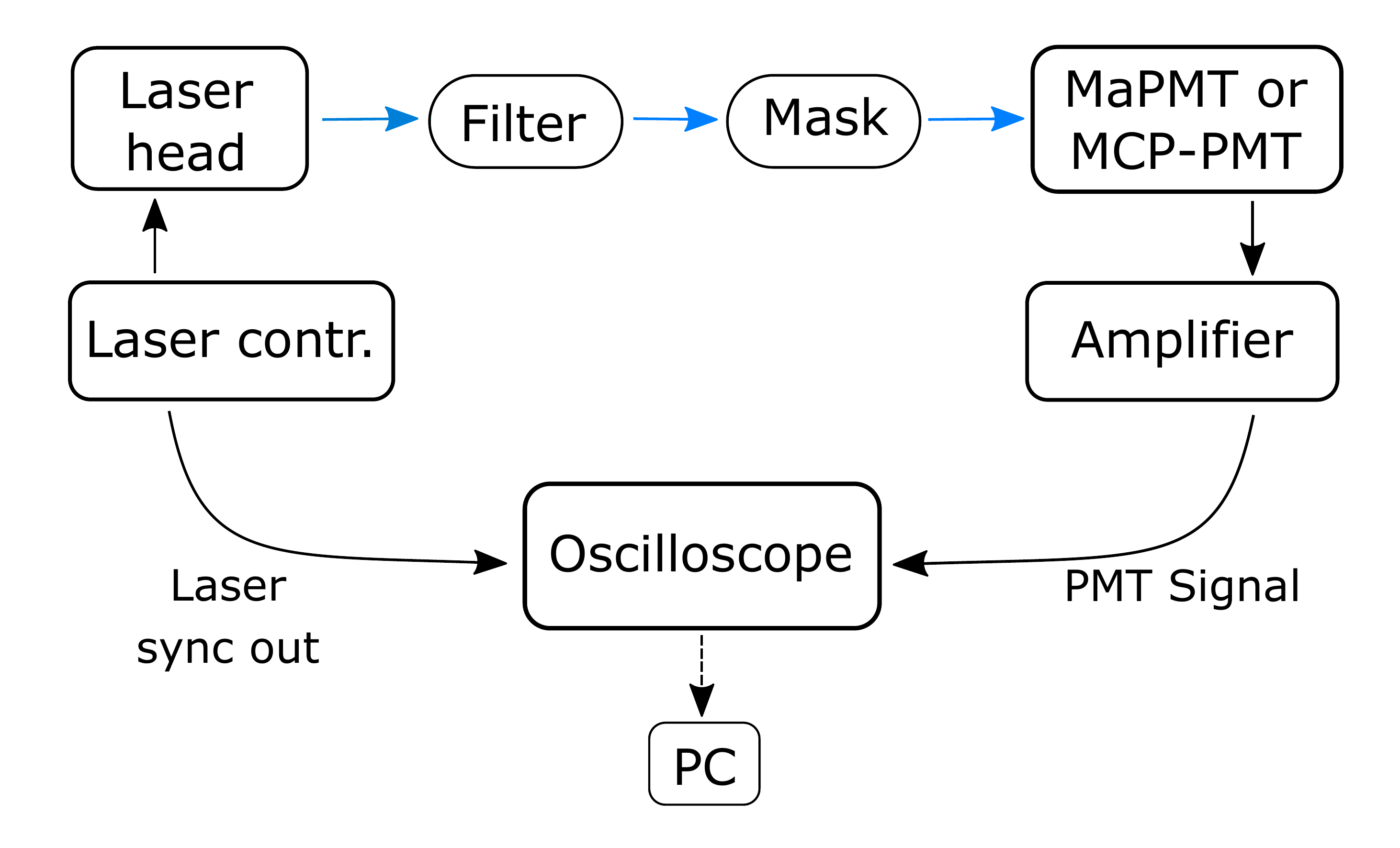}
\hspace{1cm}
\includegraphics[width=.3\textwidth]{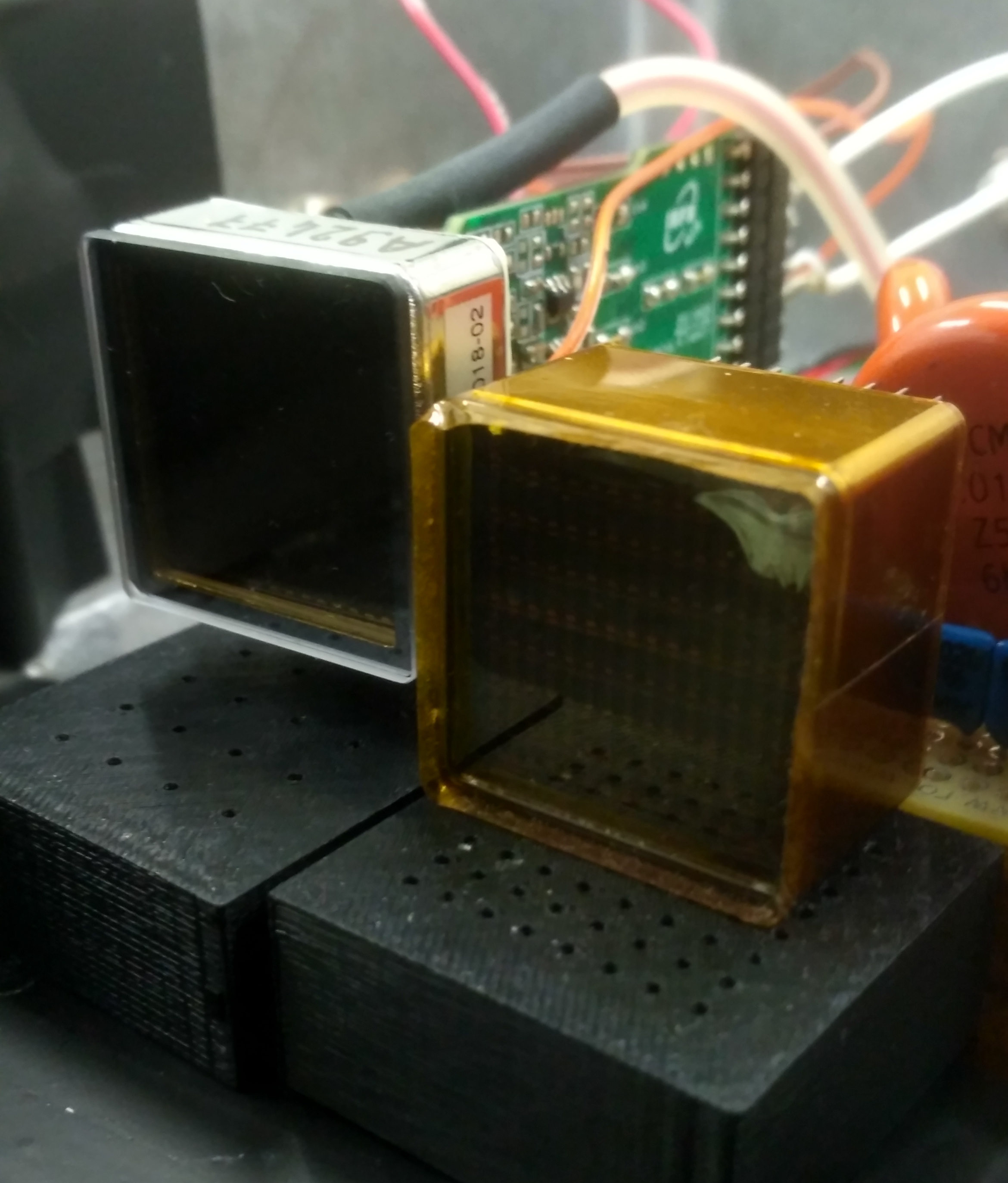}
\caption{\label{fig:testsetup} Left: block schematic of the test setup. The black arrows represent electrical signals, the blue arrows represent light. Right: photograph of the Hamamatsu R10754 MCP-PMT (left) and R13742 MaPMT (right) resting on the masks used to illuminate only the pixel centers.}
\end{figure}

Figure \ref{fig:testsetup}, left side, shows a block schematic of the setup.
The photodetector was illuminated with laser pulses of 70~ps FWHM duration and 405~nm wavelength from a Hamamatsu PLP-10 (C10196 controller and M10306-29 head).
The amplitude of the laser pulses was kept at a medium setting on the PLP-10 controller (knob set between 8 and 10). This seemed optimal, since we observed a larger pulse duration at lower settings, and a smaller secondary pulse delayed by about 200~ps at higher settings.
We placed a filter in front of the laser head to reach the single photon regime, with attenuation ranging from 10$^3$ to 10$^6$ (Thorlabs AR-coated absorptive neutral density filters NE30A-A to NE60A-A).
The filters will be denoted by F3 to F6 in the following, where the number denotes the optical density, or absorbance, of the filter.
3D-printed masks with 1~mm diameter holes were used to illuminate only the pixel centers, when needed.
Figure \ref{fig:testsetup}, right side, shows a photograph of both photodetectors.

\begin{figure}[t]
\centering
\includegraphics[trim=100px 260px 100px 250px,clip,width=.5\textwidth]{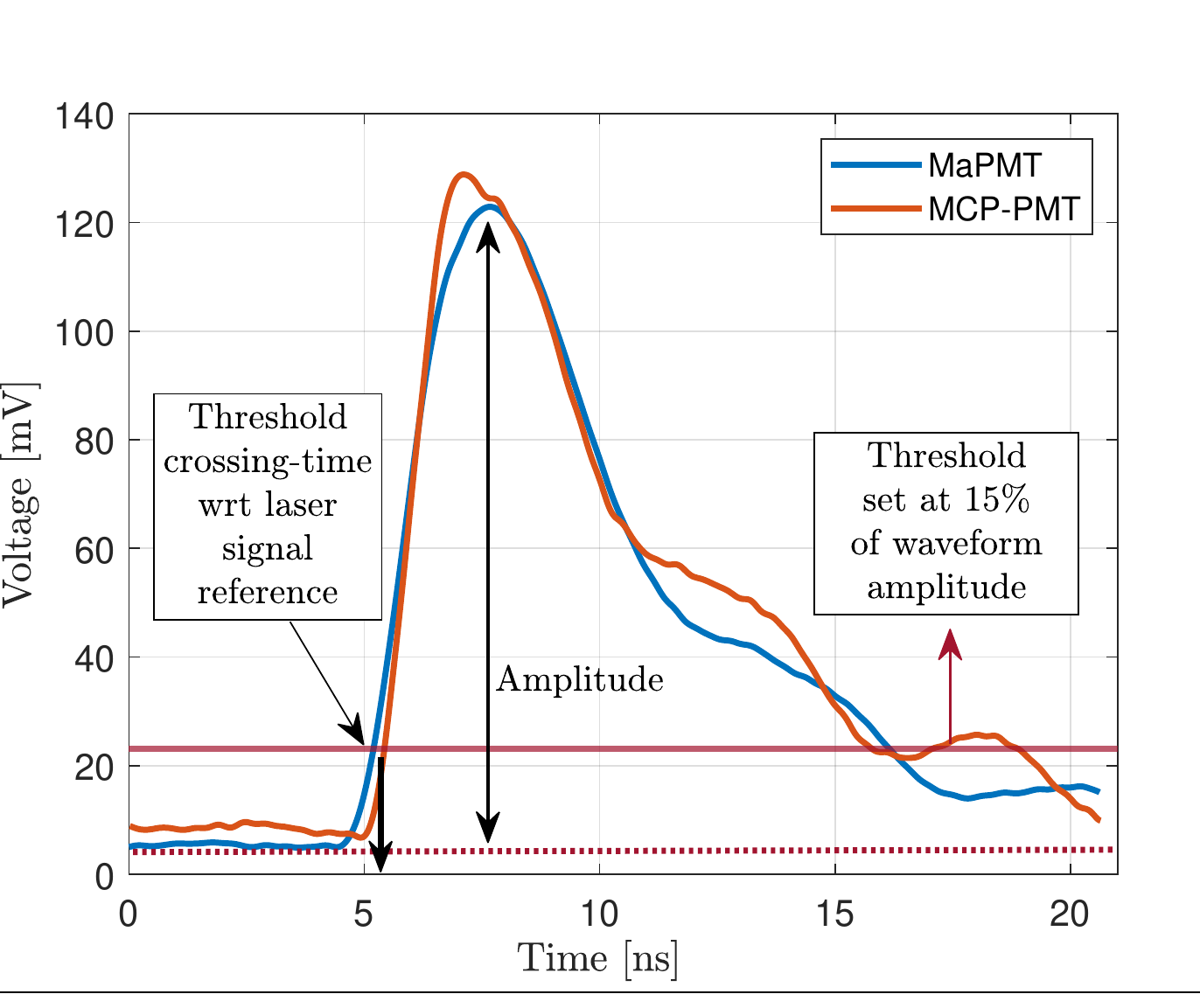}
\caption{\label{fig:onesignal} Typical signals from MaPMT and MCP-PMT, selected to have the same amplitude. They are almost identical, since they are shaped by the same amplifier. Time is detected at 15\% of the pulse amplitude.}
\end{figure}

The anode signals were read out with a LMH6702 current feedback operational amplifier, operating as a fast integrator with $C_F\simeq3$~pF (internal) and  $R_F=1$~k$\Omega$ \cite{CFopamp}.
The signals at the output of the amplifier had a rise time $t_r= $~1.5~ns and a fall time $t_f\simeq$~10~ns.
They were fed to the oscilloscope Rohde~\&~Schwarz RTO1044 (4~GHz, 20~GS/s), digitally low-pass filtered at 300~MHz and acquired.
A signal from the MaPMT and one from the MCP-PMT, selected to have the same amplitude, are shown in figure \ref{fig:onesignal}.
The measured gain, halved by the 50~$\Omega$ termination, was 27.5~mV/Me$^-$, calibrated by injecting a known charge trough a test capacitor.
The baseline noise as seen at the oscilloscope was 0.25~mV RMS, or $\sigma_Q =9$~ke$^-$ RMS, the main contributor being the current noise at the inverting input of the operational amplifier (18.5~pA/$\sqrt{Hz}$).
The effect of electronic noise on the measurement of the threshold crossing time can be estimated as $\sigma_t = \sigma_V / f'(t)$, where $\sigma_V$ is the voltage noise and $f'(t)$ is the time derivative of the signal. It can equivalently be expressed as $\sigma_t = t_r \sigma_Q / Q$, where $t_r$ is the rise time, $\sigma_Q$ the equivalent noise charge and $Q$ is the charge carried by the signal.
For signals above 10$^6$~e$^-$ (single photons at nominal PMT gain), this gives less than 15~ps RMS, negligible compared to the timing resolution of the PMT and laser.
Pixels that were not read out were connected to ground.

\begin{figure}[t]
\centering
\includegraphics[trim=100px 240px 100px 250px,clip,width=.55\textwidth]{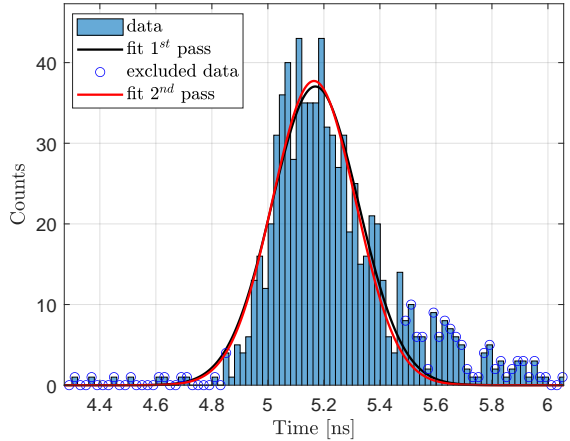}
\caption{\label{fig:histogram} Typical distribution of threshold crossing times, fitted with a Gaussian curve in two steps. The blue markers indicate data excluded from the second fit. The difference between the two fitting curves is small, indicating that the right tail gives in any case a negligible contribution.}
\end{figure}

The photon rate hitting the detector was varied by changing the laser pulse repetition rate $\nu_L$ from a few Hz up to 100~MHz.
The filters in front of the photodetector were chosen so that the actual rate of non-empty events $\nu_P$ was always below $\nu_L/10$.
Assuming the number of photons in each event to be Poisson-distributed with average $\mu$, the probability of observing $n$ photons is $P(n) = {\mu^n e^{-\mu}}/{n!}$.
The probability of observing a non-empty event is $ 1-P(0) = 1- e^{- \mu} \simeq \mu$, where the approximation is valid if $\mu$ is small. This is also equal to $\nu_P / \nu_L$.
The fraction of non-empty events that contain two or more photons is given by
\begin{equation}
    \frac{1-P(0)-P(1)}{1-P(0)} = \frac{1- e^{- \mu}-\mu e^{- \mu}}{1- e^{- \mu}}
    \simeq \frac{\mu^2/2}{\mu} = \frac{\mu}{2}.
    \label{eq:Poisson}
\end{equation}
By choosing the filter to get $\nu_P < \nu_L/10$, or $\mu < 0.1$, according to \ref{eq:Poisson} the non-empty signals were at $>95\%$ single photons, which ensures a sufficient purity of single photon events.
To avoid acquiring lots of empty signals, the oscilloscope was set to trigger on the sequence (within a 100~ns window) of a signal from the photodetector, detected with a threshold just above noise, and the delayed by 20~ns ``sync~out'' signal from the laser controller. The latter provided the time reference for each signal.
For low and medium rate measurements, all signals were acquired by the oscilloscope.
For the highest rate measurements, where $\nu_P$ exceeded the maximum trigger rate capability of the oscilloscope (a few MHz), the oscilloscope could only acquire a fraction of the total.
The actual photon rate in these measurements was estimated by replacing the filter with a $10 \times$ or $100 \times$ higher attenuation, measuring the rate in the same conditions except for the filter, and then scaling it back by multiplying it by 10 or 100.
The validity of this method was verified by comparing the measured photon rates with F3 and $\nu_L=1$~kHz, F4 and $\nu_L=10$~kHz, F5 and $\nu_L=100$~kHz, observing similar values of $\nu_P$.

An offline algorithm was used to analyze the acquired waveforms.
A threshold at 15\% of its amplitude was applied to each signal, as shown in figure \ref{fig:onesignal}.
This is equivalent to constant fraction discrimination, and avoids spurious contributions to time resolution due to amplitude walk.
The threshold crossing times of a given set of waveforms were collected in a histogram binned at 20~ps.
An example is shown in figure \ref{fig:histogram}.
There is an asymmetric tail at the right side of the histogram (delayed signals).
This is likely due to recoil electrons inside the MCP, although some contribution from the laser cannot be excluded.
The Gaussian fit was performed in two steps: first the entire distribution was fitted, to determine mean and standard deviation $\sigma$ (1st pass). The points that lie more than $\pm 2 \sigma$ away from the mean of the distribution were then excluded, and the remaining data was fitted again (2nd pass).
The difference between the two fitting Gaussian curves is anyway often negligible, as can be seen in figure \ref{fig:histogram}, where the two curves are nearly identical.
We checked the stability of the algorithm by changing the threshold and the bin width around these values, and did not observe significant variations in the results.
The error bars associated with each measurement represent the statistical uncertainty on the fit parameters ($1 \sigma$ confidence level).

\section{Results}

\begin{figure}[t]
\centering
\includegraphics[trim=100px 240px 100px 250px,clip,width=.7\textwidth]{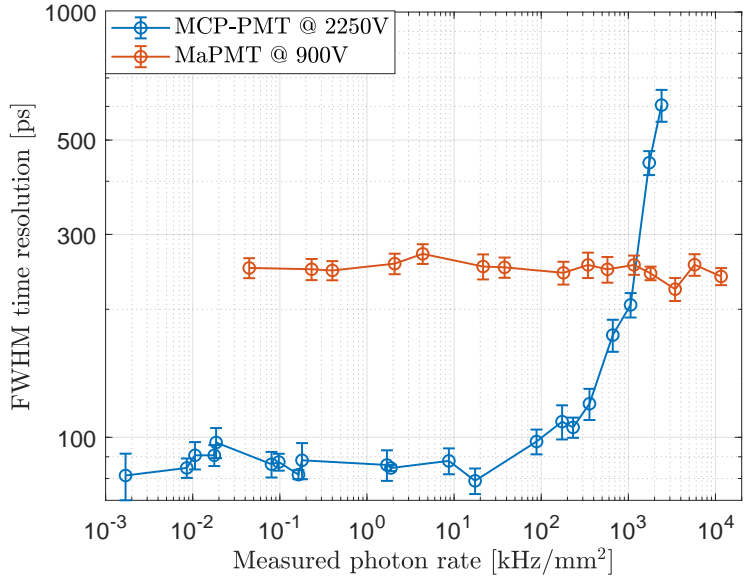}
\caption{\label{fig:timingVSrate} Measured timing resolution for MaPMT and MCP-PMT as a function of rate.}
\end{figure}

Figure \ref{fig:timingVSrate} shows the measured timing resolution of the MaPMT and the MCP-PMT as a function of rate.
A mask was placed in front of the devices, illuminating just the pixel centers with holes of 1~mm diameter.
The measured photon rate $\nu_P$ was scaled per mm$^2$ using the hole area (0.785~mm$^2$).
At the operating voltages noted in the plot legend, gain was above 10$^6$ for both devices. In this gain range the noise of the readout chain was negligible, and the only contributions to time resolution are from the photodetector and the laser.
The time resolution of the MaPMT in these conditions is 250~ps~FWHM, independent of rate up to 10~MHz/mm$^2$, the highest tested.
Measurements of the MCP-PMT at low rate give a time resolution of 90~ps~FWHM, about equally contributed by the laser pulse duration ($\leq 70$~ps~FWHM) and the MCP-PMT ($\sim$70~ps~FWHM) summed in quadrature.
At rates above 100~kHz/mm$^2$ the resolution of the MCP-PMT begins to degrade. The curve rapidly points upward, overtaking that of the MaPMT at about 1~MHz/mm$^2$.
This is due to saturation of the MCP. It happens when the average interval between photon signals hitting the same microchannel becomes comparable or smaller than the time it takes to recharge the channel walls after a signal.

\begin{figure}[t]
\centering
\includegraphics[trim=100px 270px 100px 250px,clip,width=.6\textwidth]{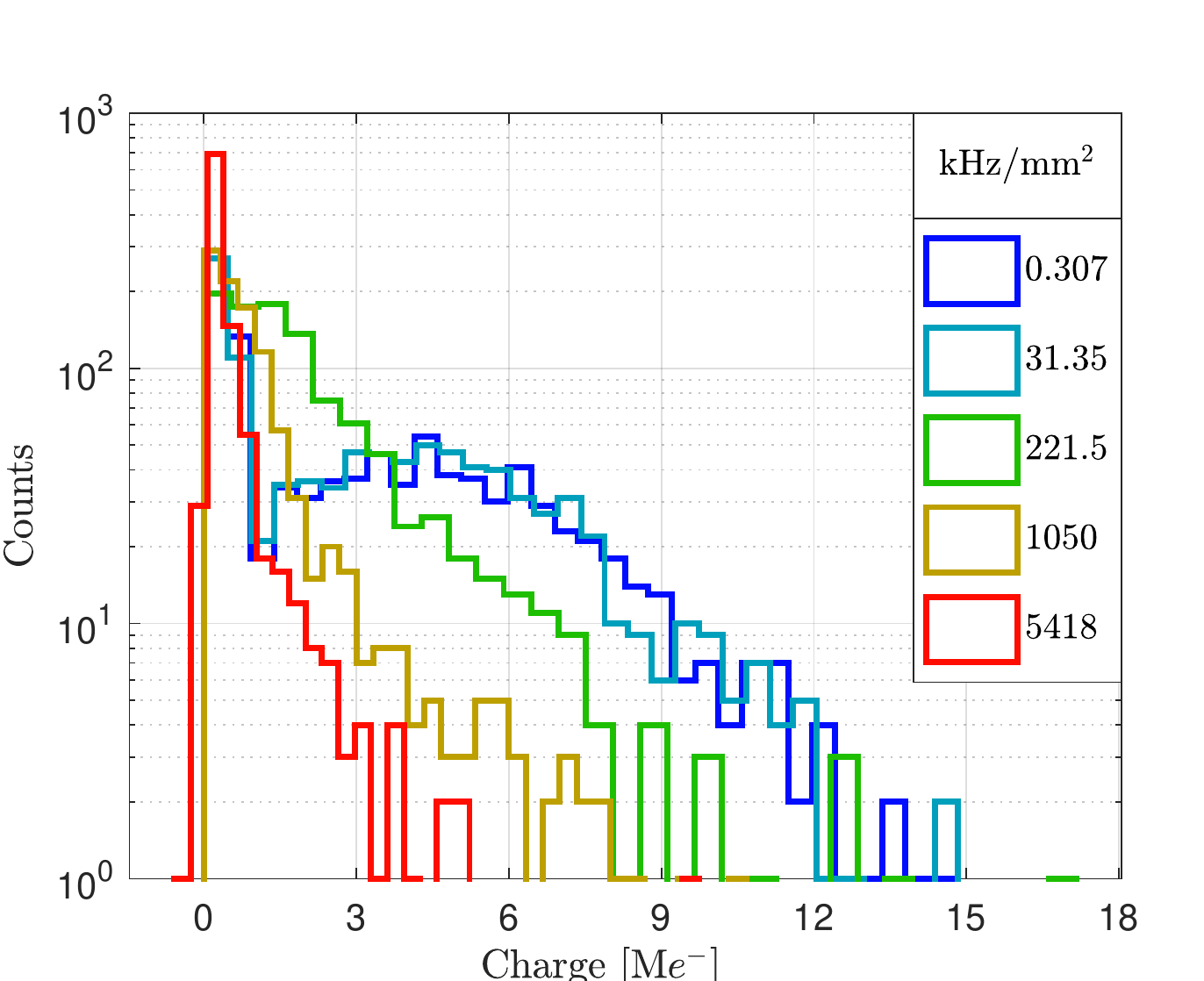}
\caption{\label{fig:MCPspectra} Single photon spectra of a pixel of the MCP-PMT at different photon rates.}
\end{figure}

\begin{figure}[t]
\centering
\includegraphics[trim=100px 240px 100px 250px,clip,width=.7\textwidth]{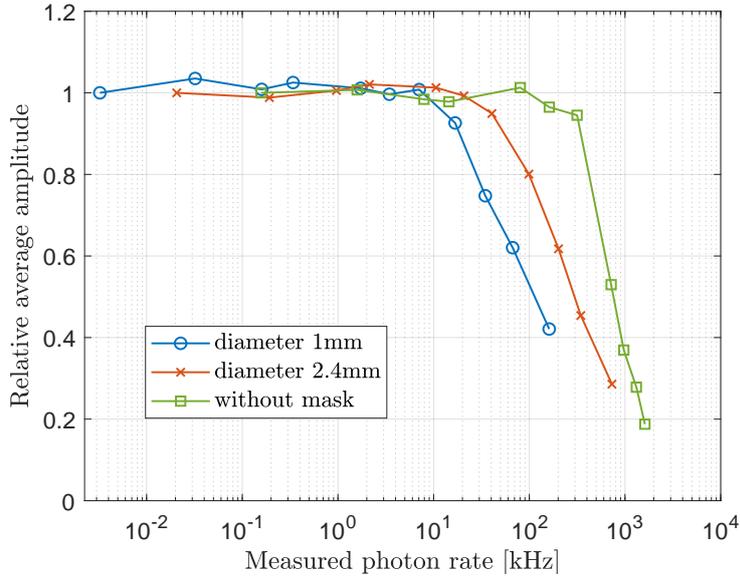}
\caption{\label{fig:MCPsaturationmask} Average amplitude of MCP signals as a function of photon rate, when different pixel areas are illuminated. The vertical scale is relative to the point at the lowest rate for each curve.}
\end{figure}

Figure \ref{fig:MCPspectra} shows the single photon spectra of the MCP at different rates.
Up to 30~kHz/mm$^2$ the single photon peak is still separated from the noise pedestal, and the spectrum is indistinguishable from those taken at a much lower rate. At higher rate saturation occurs. The spectra at 220~kHz/mm$^2$ and above show that the average gain is strongly reduced. Since this is due to the increasing number of microchannels that are not fully recharged when they are hit by the next signals, the reduction of average gain goes together with loss of efficiency.

Figure \ref{fig:MCPsaturationmask} shows the average amplitude of the anode signals as a function of absolute signal rate, with increasing illuminated area on the pixel.
The amplitude is normalised to the value at the lowest rate.
The first effects of saturation are visible above $\sim$10~kHz with the 1~mm diameter hole (0.785~mm$^2$), $\sim$50~kHz with the 2.4~mm diameter hole (4.52~mm$^2$), and $\sim$400~kHz when the entire pixel area ($5.28 \times 5.28$~mm$^2=27.9$mm$^2$) is uncovered.
The ratio of saturation rate and illuminated area (proportional to the number of microchannels) is approximately constant, confirming that it is a local phenomenon.
Going back to figure \ref{fig:timingVSrate}, it is anyway interesting to note that the time resolution is stable below 100~ps~FWHM up to the onset of saturation, and even slightly beyond.

The results on time resolution at high signal rate are compatible with those in \cite{MCPBelle} for devices of the same family, independently of MCP resistance. Concerning gain reduction, our results are compatible with those in \cite{MCPBelle} for high resistance MCPs. In that work, MCPs of smaller resistance were shown to withstand illumination rates beyond $\sim$2~MHz/anode, or $\sim$100~kHz/mm$^2$, without gain loss.
Our results on rate stability are also compatible with those presented in \cite{MCPPanda} for a 2-inch device based on the same technology (Hamamatsu R13266).

\begin{figure}[t]
\centering
\includegraphics[trim=100px 240px 100px 250px,clip,width=.7\textwidth]{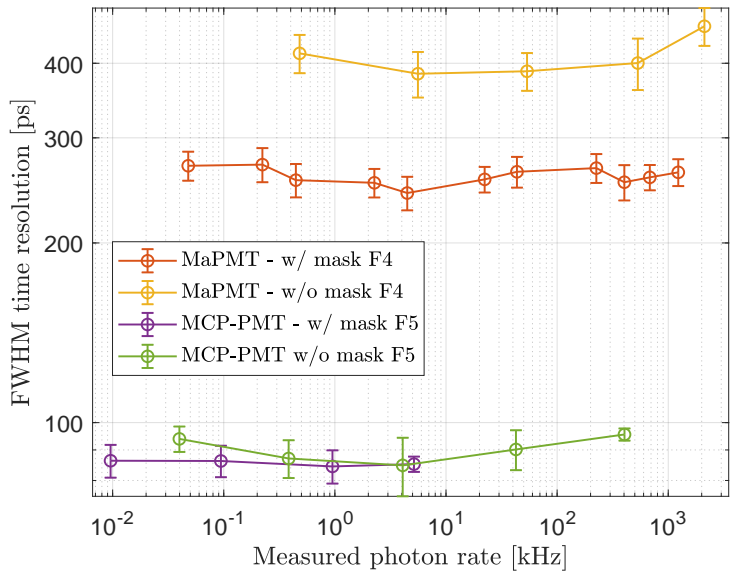}
\caption{\label{fig:withwithoutmask} Time resolution of the MaPMT and MCP-PMT with and without the mask used to illuminate only the pixel center. Removing the mask has no effects on the MCP-PMT, while the resolution of the MaPMT degrades to 400~ps FWHM.}
\end{figure}

Besides scaling the absolute pixel rate, removing the mask has no effect on the time resolution of the MCP-PMT. This is not the case for the MaPMT.
Figure \ref{fig:withwithoutmask} compares measurements taken with and without the mask in front of the photodetectors.
When the mask is removed, the resolution of the MaPMT degrades from 250~ps~FWHM to 400~ps~FWHM, independently of rate.
This was investigated by offsetting the mask hole to illuminate only the pixel boundary instead of the center. Photons hitting the pixel boundary region were observed to have both a longer transit time (arriving late by 300~ps on average) and a larger transit time spread ($\sim$440~ps FWHM) compared with photons hitting the pixel center.
When the entire pixel area is uncovered, this leads to an overall resolution of 400~ps FWHM as measured and shown in figure \ref{fig:withwithoutmask}.

\section{Conclusions and outlook}

We presented a comparison of the timing performance of a multi-anode PMT (Hamamatsu R13742) and a MCP-PMT (Hamamatsu R10754).
The MaPMT offers stable operation up to 10~MHz/mm$^2$. Despite the fact that illuminating only the pixel center gives a timing resolution of 250~ps FWHM, its resolution when the entire active surface is illuminated is 400~ps FWHM, setting it far apart from the timing performance of the MCP-PMT.
The MCP-PMT offers superior timing resolution compared to the MaPMT, but the maximum sustainable rate before saturation at nominal gain is just below 100~kHz/mm$^2$.
Operation of MCP-based photodetectors as single photon counters at  MHz/mm$^2$ rates and above will likely require using microchannels of smaller diameter and lower resistivity, to reduce the hit probability and recharge time of each microchannel. These come with stimulating, but currently unsurmounted, technological challenges for MCP manufacturing, cooling and operation.

\end{document}